\documentclass[]{spie}  

\usepackage{amsmath,amsfonts,amssymb}
\usepackage{graphicx}
\usepackage[colorlinks=true, allcolors=blue]{hyperref}

\title{In-flight performance and calibration of the Grating Wheel Assembly sensors (NIRSpec/JWST)}

\author[a]{Catarina Alves de Oliveira}
\author[b]{Nora L\"{u}tzgendorf}
\author[c]{Peter Zeidler}
\author[d]{Giovanna Giardino}
\author[a]{Pierre Ferruit}
\author[e]{Nimisha Kumari}
\author[b]{Timothy Rawle}
\author[b]{Stephan M. Birkmann}
\author[b]{Torsten B\"{o}ker}
\author[e]{Charles Proffitt}
\author[b]{Marco Sirianni}
\author[b]{Maurice Te Plate}

\affil[a]{European Space Agency, ESAC, Madrid, Spain}
\affil[b]{European Space Agency, STScI, Baltimore, USA}
\affil[c]{AURA for the European Space Agency, STScI, Baltimore, USA}
\affil[d]{ATG Europe for the European Space Agency, ESTEC, The Netherlands}
\affil[e]{Space Telescope Science Institute, Baltimore, USA}

\authorinfo{Further author information: (Send correspondence to C.A.d.O.)\\C.A.d.O.: E-mail: catarina.alves@esa.int}

\begin{document} 
\maketitle

\begin{abstract}

The Near-Infrared Spectrograph (NIRSpec) on board of the James Webb Space Telescope will be the first multi-object spectrograph in space offering $\sim$250,000 configurable micro-shutters, apart from being equipped with an integral field unit and fixed slits. At its heart, the NIRSpec grating wheel assembly is a cryogenic mechanism equipped with six dispersion gratings, a prism, and a mirror. The finite angular positioning repeatability of the wheel causes small but measurable displacements of the light beam on the focal plane, precluding a static solution to predict the light-path. To address that, two magneto-resistive position sensors are used to measure the tip and tilt displacement of the selected GWA element each time the wheel is rotated. The calibration of these sensors is a crucial component of the model-based approach used for NIRSpec for calibration, spectral extraction, and target placement in the micro-shutters. In this paper, we present the results of the evolution of the GWA sensors performance and calibration from ground to space environments.
\end{abstract}


\section{INTRODUCTION}
\label{sec:intro}  

The Near-Infrared-Spectrograph (NIRSpec)\cite{Jakobsen2022} is one of the science instruments aboard the James Webb Space Telescope (JWST)\cite{gardner2006} launched on 25 December 2021. NIRSpec is a versatile instrument enabling low-, medium-, and high-resolution ($R{\simeq}100$, 1000, and 2700, respectively) near-infrared spectroscopy in support of all JWST science themes. To that end, it offers four different observing modes: 1) integral field spectroscopy (IFS)\cite{Boeker2022}, 2) multi-object spectroscopy (MOS) via a micro-shutter array that employs about 250,000 individually addressable shutters\cite{Ferruit2022} , 3) fixed slit (FS) spectroscopy via several long slits, and 4) bright object times series (BOTS)\cite{Birkmann2022} observations via a dedicated square aperture. 

These proceedings address the in-flight performance and calibration of the NIRSpec Grating Wheel Assembly\cite{Weidlich2006} (GWA), a cryogenic wheel mechanism that can be configured to position one of its optical elements into the beam path. It is equipped with six dispersion gratings ($R{\simeq}1000$
and $R{\simeq}2700$), a a double pass-prism ($R{\simeq}100$), and a mirror for target acquisition. The rotational degree of freedom of the wheel is given by a ball bearing controlled by two mechanisms: a cryogenic torque motor used as actuator, and a spring operated ratchet to achieve accurate positioning. Additional electrical components include temperature and tip/tilt sensors, and a harness that connects them to the unit\cite{Weidlich2008}.

The GWA components are located in the pupil plane of the instrument, dispersing or reflecting the beam to the camera that focuses it onto the focal plane assembly. The optical alignment of the wheel at instrument level, and that of the optical elements with respect to each other, ensures maximum throughput and minimum stray-light and minimizes any image displacement in the instrument’s field of view. However, the ball bearing mechanism limits the mechanical angular reproducibility of the wheel, resulting in a bore-sight shift of the selected optical element every time the wheel is moved. Additionally, any remaining tilt of the bearing axis with respect to the bearing flange will also cause a deviation of the beam path (known as ‘wobbling’). Combined, these effects produce a displacement of the beam reflected from the optical element on the focal plane in both the spatial and spectral direction, impacting the position of the dispersed or reflected beam on the focal plane. This effect impacts not only the analysis of science data, but also the instrument operations. The NIRSpec target acquisition procedure will rely on a set of on-board software operations to derive any necessary correction of the telescope pointing on sky as to ensure the science targets are acquired at the desired aperture\cite{Keyes2018}. In practice, the telescope pointing correction is given by the difference between the known target sky coordinates and those measured on the focal plane array (FPA) and transformed to the sky plane. The necessary geometrical transforms for this later computation are defined by the NIRSpec instrument model\cite{LuetzgendorfSPIE}. The position of the grating wheel at the moment the target acquisition image is acquired defines the position of the reflected beam on the focal plane and therefore it must be known to a great accuracy in this computation. NIRSpec’s stringent operation mode requires the angular position of the GWA mechanism to be known to a higher accuracy, i.e., approximately 5 and 10 times better than the typical reproducibility for spectral calibration and target acquisition, respectively. To overcome the limitations on the accuracy to which the grating wheel can be positioned, two position sensors are used to measure the tip/tilt pointing error of each selected GWA optical element every time the wheel is moved\cite{Marchi2012}. Here we present the in-orbit calibration of the GWA position sensors which was carried out during the JWST Commissioning Phase \cite{BoekerSPIE}. 

\section{DATA}

The GWA calibration presented in these proceedings is based on data acquired during the the in-flight Commissioning (COMM) period from launch on 25 December 2021 until the final NIRSpec commissioning activity on 28 June 2022\cite{BoekerSPIE,Rigby2022}. 

The data were acquired using internal lamps which are part of the calibration assembly unit (CAA) inside the instrument, and provide a spatially uniform illumination of the NIRSpec slit plane. For the analysis of the sensors when using the mirror element, imaging data was collected by acquiring exposures of the MSA illuminated by the TEST lamp, configured in a checkerboard pattern, where a subset of individual shutters can be easily analysed. To characterize the behaviour of the sensors for the dispersers, spectral references provided by sources in the CAA were used to illuminate the fixed slits and the MSA.

The exposures were first processed with the NIRSpec Commissioning Team ramps-to-slopes pipeline that applies the following corrections: bias subtraction, reference pixel subtraction, linearity correction, dark subtraction and finally count-rate estimation, including jump detection and cosmic-ray rejection\cite{Birkmann22b}. Additionally, for the spectral data, from the count-rate images the wavelength calibrated spectra were obtained using the second part of the NIRSpec calibration pipeline\cite{AlvesdeOliveira2018} to perform the following operations for the fixed slit: extract sub-image containing the spectral trace and assign wavelength and spatial coordinates to each pixel therein; generate a rectified spectrum re-sampled on regular 2D-grid; compute the 1D-spectrum obtained by spatial integration. One of the corrections implemented by the pipeline is the position of the grating wheel at each configuration, precisely the parameter for which we want to improve the calibration. Therefore, when extracting the spectra we disabled this correction so that the spectra are extracted assuming that for each configuration the respective disperser is exactly at the nominal position. In total, over 300 exposures were used in the analysis.

\section{METHODS}
\label{sec:method}

The first step of the calibration is to derive for each exposure the location of the wheel. For that, we make use of the NIRSpec spectrograph model derived during Commissioning \cite{LuetzgendorfSPIE}. The model has been extensively described in the literature\cite{Dorner2016,Giardino2016}, and, in summary, has two major components: (1) the parameterization of the coordinate transforms between the main optical planes (IFU-FORE, IFU-POST, collimator, and camera); and (2) the geometrical description of their elements (MSA, IFU slicer, GWA, FPA). In particular, the GWA elements are described by their orientational positioning in the wheel, and according to type, with the gratings defined by the individual groove densities and front surface tilt angles, the prism by the front surface tilt, internal prism angle, and the Seidel model based prescription for its refractive index, and the mirror treated as a simple reflective surface.

For each exposure, a reduced fitting procedure of the NIRSpec instrument model is performed. The reference points to use in the model fit are the derived centroids of the shutters in the imaging data, for which the package Source Extractor\cite{Bertin1996} is used, or the centroids of absorption and emission spectral features which are derived using the center of gravity method. To constrain the spatial location of the spectra, a polynomial fit to the continuum is performed. The model fit follows then an iterative approach. First, a manual adjustment is done, where large differences between the predicted and observed image and spectra location on the detectors are removed through an approximate derivation of the correct GWA angles that define the wheel position. Secondly, taking these angles as the initial condition, and freezing all other model parameters, the two angles defining the wheel position are set as free parameters and are simultaneously modified by a least square fit to find the best solution by minimizing the differences between the model predicted and observed positions of the centroids of shutters in the imaging data or line centroids in the spectral data.

Finally, the calibration of a given sensor is then derived by establishing the linear relation between the telemetry reading from the sensor and the derived location of the wheel for the corresponding angle.

\section{RESULTS}
 
\subsection{Determining the location of the wheel}
\begin{figure}
\centering
  \includegraphics[height=8cm]{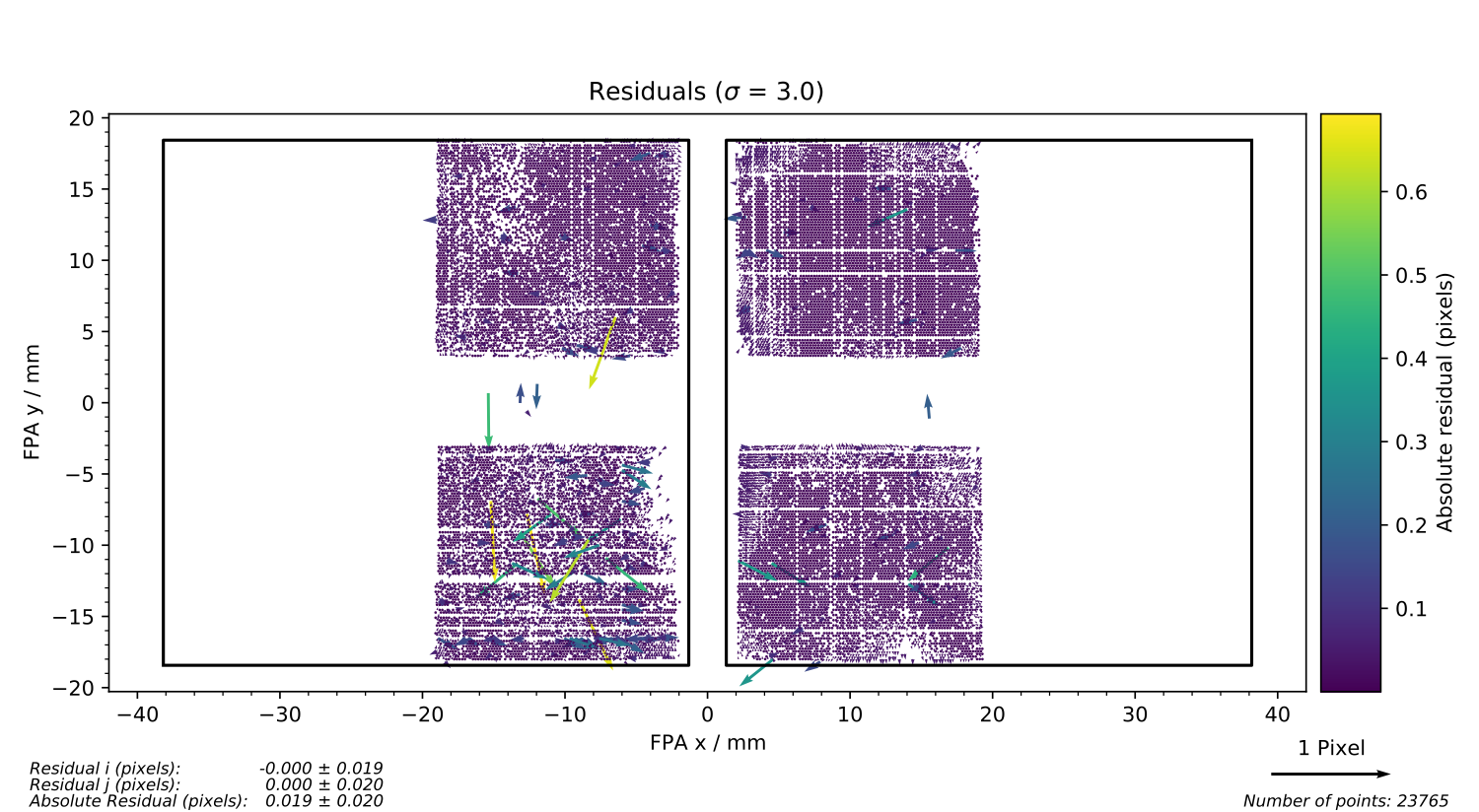}
    \caption[example] {Example of the residuals obtained for a single imaging exposure calculated as the difference between the model predicted and measured shutter centroids at the detectors for the best-fit GWA position for that exposure.}
    \label{fig:fit}
\end{figure}

For each of the more than 300 exposures, the location of the wheel was determined using the method described above. The residuals of the least square fit were confirmed to be below the required level of 0.05 pixels has measured on the detector plane for every exposure. Figure~\ref{fig:fit} shows an example of the residuals for a single imaging exposure calculated as the difference between the model predicted and measured shutter centroids at the detectors for the best-fit GWA position for that exposure.

\subsection{Calibration relation}
\begin{figure}
\centering
\begin{tabular}{l c} 
  \includegraphics[height=9cm]{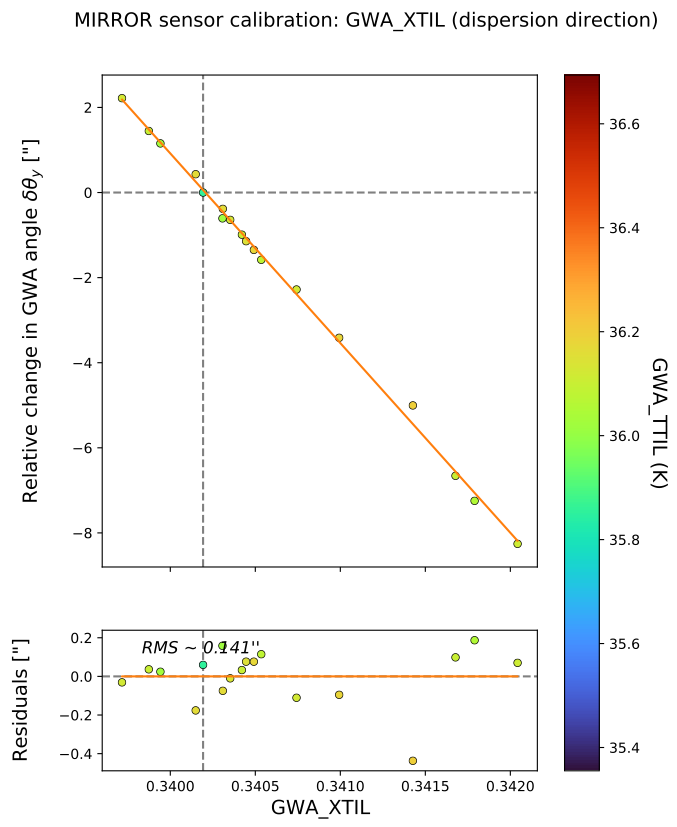} &
  \includegraphics[height=9cm]{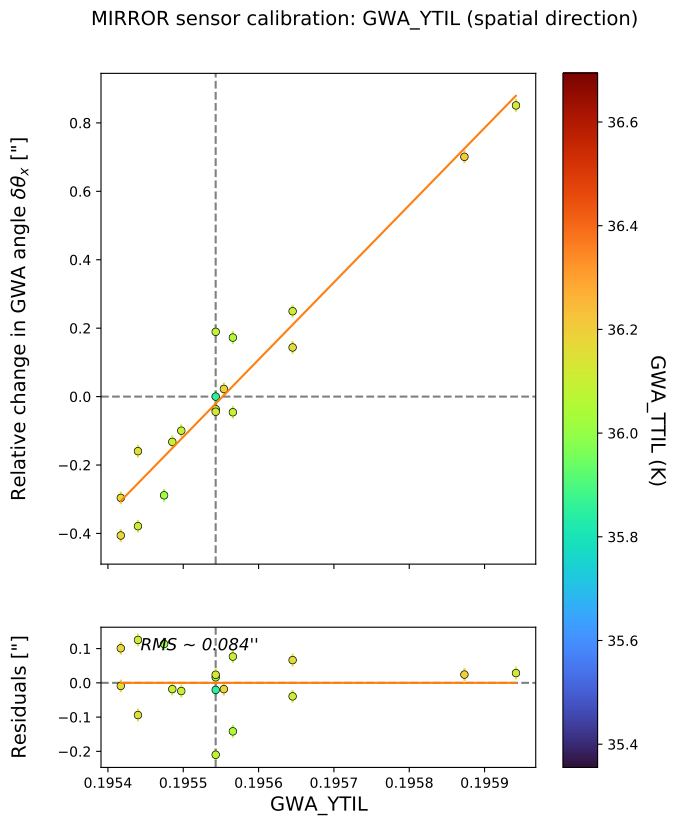} \\
  \includegraphics[height=9cm]{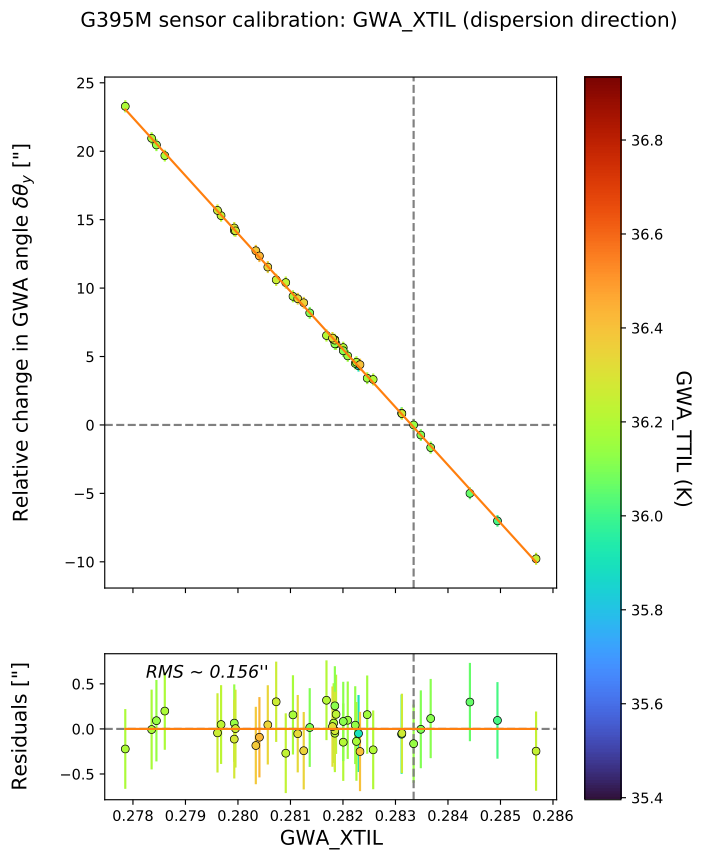} & \includegraphics[height=9cm]{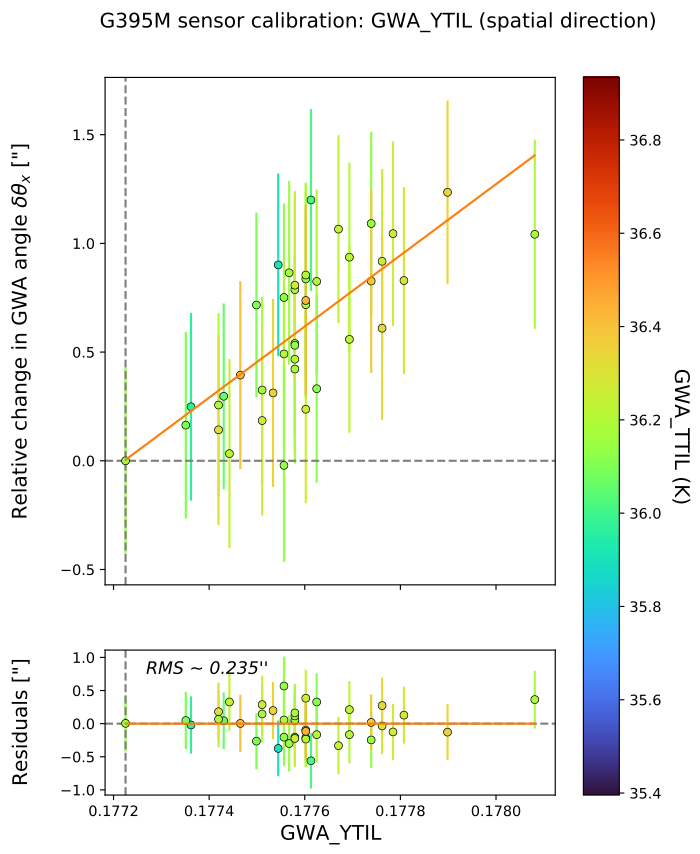} \\
\end{tabular}
\caption[example] {Calibration relations using the derived GWA position for each exposure, and the respective sensor readings for the mirror (top panels) and one example grating, G395M (lower panels). The colorbar represents the temperature of the wheel for each exposure as measured by a dedicated sensor.}
      \label{fig:calib}
\end{figure}

For each instrumental set-up, we compare the telemetry from the sensor voltage reading to the angular shift of the wheel derived through the reduced model fit technique. Example results are presented in Figure~\ref{fig:calib}, showing the linear fit to the data and the fit to the residuals for the mirror and one of the gratings, respectively. These relations are known to change with the temperature of the optical bench, and therefore, and as expected, the values computed differ from the ones previously derived in ground-test campaigns\cite{AlvesdeOliveira2018}. Since the calibration has been done at the instrument final operating temperature, these represent the in-orbit relations for science operations.

The R.M.S. of the residuals of the linear calibration provides an estimate of the accuracy of the sensor calibration method presented here. For all optical elements and sensor readings, the behaviour of the sensors has remained extremely stable from the ground to the in-orbit performances, and the calibration between the sensor readings and the derived GWA angular displacement is shown  to be well represented by a linear relation.

\subsection{Applying the calibration relation}

\begin{figure}
\centering
\includegraphics[height=8cm]{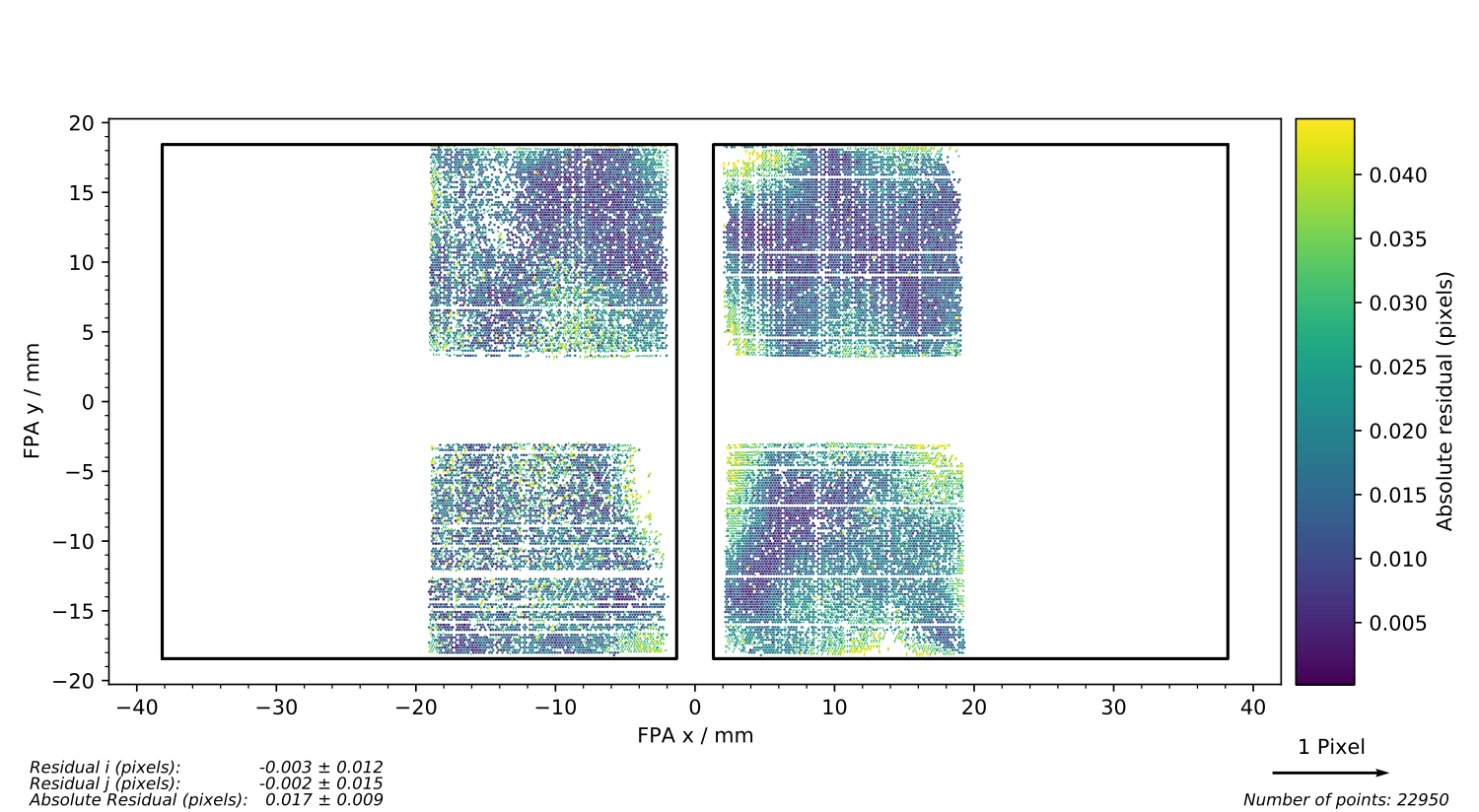} \\
\caption{Residuals showing the difference between model predicted and measured shutter centroids at the detectors using the in-orbit derived GWA sensor calibration for the mirror.}
  \label{fig:verifymirror}
\end{figure}

\begin{figure}
\centering
\includegraphics[height=8cm]{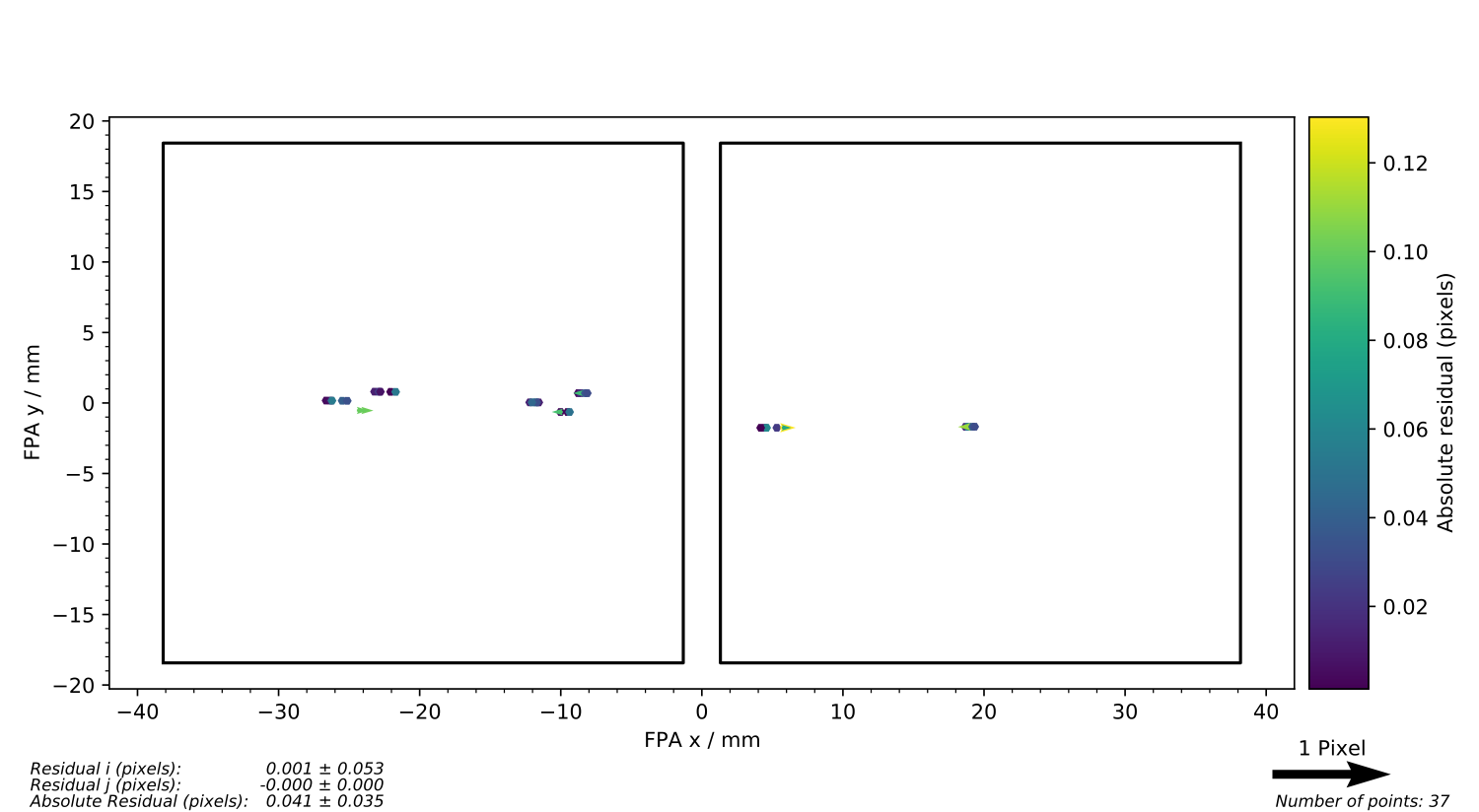}
\caption[example] {Residuals showing the difference between model predicted and measured spectral line centroids measured for the NIRSpec fixed slits at the detectors using the in-orbit derived GWA sensor calibration for the G395M grating.}
    \label{fig:verifydisp}
\end{figure}

To verify the performance of the GWA calibration, all the data was then re-processed taking into account the updated linear relation coefficients to use the best predicted wheel location for each exposure. The reference points were then re-calculated, and the residuals between the new model-predicted values compared to the measurements. Some example results are shown in Figures~\ref{fig:verifymirror} and \ref{fig:verifydisp} for one imaging and one spectral exposures, respectively. The residuals are consistently below 1$/$10th of a pixel, confirming the excellent performance of the instrument model both for imaging and spectral modes\cite{LuetzgendorfSPIE}.

\section{DISCUSSION AND CONCLUSION}

The analysis and results from the calibration of the magneto-resistive position sensors installed on NIRSpec’s GWA performed in-orbit during the Commissioning phase of JWST confirms that they can be reliably used to derive the position of the wheel for every given exposure with the required accuracy. 

While it was observed during the Commissioning period that the mechanical reproducibility of the wheel leaves uncertainties of several pixels on the position of spectral features or reflected images of sources on the detector, the sensor voltage reading has sufficient accuracy to overcome that shortcoming and deliver the necessary performance of the wheel. 

In particular, the applied methodology to use a reduced instrument model fit on every single calibration exposures acquired with internal lamps to derive the wheel location was shown to work remarkably well in-orbit, and allowed for the derivation of the linear relation between the sensor voltage reading and the wheel location. 

The derived calibrations are already in use for NIRSpec nominal science operations, both by integrating the on-board target acquisition procedure, which has been successfully applied to a number of observing programs, and on the science data processing. 

In conclusion, the analysis presented in these proceedings shows that the sensors installed on NIRSpec's grating wheel can be reliably used in-orbit to determine the location of the wheel, which in turn, allows the accurate derivation of the position on the detector of any spectral feature of known wavelength with an accuracy higher than that required for wavelength calibration or any source imaged for target acquisition. 

\bibliography{report} 
\bibliographystyle{spiebib} 

\end{document}